\documentstyle[lprocl]{article}
\begin{document}
\vspace*{-2.cm}
\title{CONFINEMENT AND LIGHT-FRONT QCD
\footnote{Talk given at $3^{rd}$ AUP workshop on "Confinement, Chaos and
Collision", Paris, France, June 96}
}
\author{M.~BURKARDT}
\address{Department of Physics,New Mexico State University \\
Las Cruces, New Mexico 88003-0001 USA}
\maketitle
\begin{abstract}
Numerical results for the (rest-frame)
$Q\bar{Q}$ potential in  light-front quantized
$QCD_{2+1}$ on a $\perp$ lattice are presented.
Both in the longitudinal as well as the $\perp$ spatial
directions one obtains linear confinement. The confinement
mechanism in light-front QCD depends on the orientation of the
external charges: for longitudinally (with respect to the boost
direction in the infinite momentum frame) separated quarks, confinement
arises from the instantaneous interaction piece in the Hamiltonian 
while for $\perp$ separated quarks, the confinement mechanism is similar to the
one in Hamiltonian lattice QCD. Nevertheless, already a very simple
ansatz for the effective link-field potential yields an almost 
rotationally symmetric $Q\bar{Q}$ potential. The momentum carried by the glue
strongly depends on the orientation ("polarization") of the
$Q\bar{Q}$ pair.
\end{abstract}
\begin{figure}
\unitlength.45cm
\begin{picture}(15,9.3)(-7,5.5)
\put(1.5,8.5){\line(0,1){1.7}}
\put(1.,11.1){\makebox(0,0){(discrete)}}
\put(1.,12.0){\makebox(0,0){$\perp$ space}}
\put(1.5,12.6){\vector(0,1){1.6}}
\put(2.,8.){\line(3,-1){1.2}}
\put(4.5,6.9){\makebox(0,0){long. space}}
\put(4.5,6.2){\makebox(0,0){(continuous)}}
\put(7.1,6.3){\vector(3,-1){1.8}}
\put(10.4,5.8){\line(3,1){1.8}}
\put(14.5,6.2){\makebox(0,0){(continuous)}}
\put(14.5,6.9){\makebox(0,0){time}}
\put(15.8,7.6){\vector(3,1){1.8}}
\includegraphics{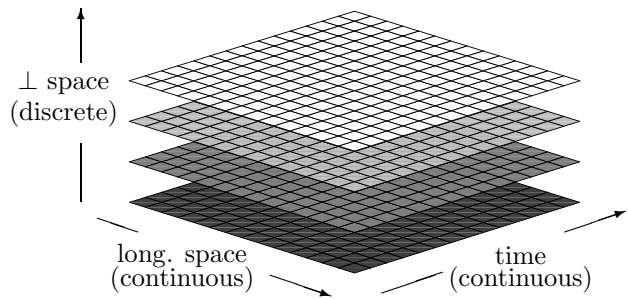}
\end{picture}
\caption{Space time view of a transverse lattice}
\label{fig:perpl}
\end{figure}

Light-front (LF) quantization is the most physical approach to
calculating parton distributions on the basis of QCD~\cite{adv}.
In the transverse lattice formulation of QCD~\cite{bardeen}, 
one keeps the
time direction and one spatial direction continuous and discretizes
the transverse spatial directions. The space time geometry is thus
an array of $1+1$ dimensional sheets (Fig. \ref{fig:perpl}).
One the one hand, this provide
both UV and IV cutoffs in the transverse directions and on the
other hand one can still perform LF quantization since the
longitudinal directions are continuous. Furthermore, in the
compact formulation, it is straightforward to implement Gauss'
law as a constraint on the states --- avoiding troublesome
divergences for $k^+\rightarrow 0$, which plague many other formulations
of LF QCD.

Another advantage of the transverse lattice is that confinement
is manifest in the limit of large $\perp$ lattice spacing $a_\perp$.
The mechanism differs for the longitudinal and the $\perp$
directions: if one separates a $Q\bar{Q}$ pair longitudinally
then, since $a_\perp$ is large, the fields in different sheets
couple only weakly and the quarks interact only with fields
in the same sheet, i.e. effectively the theory reduces to 
$1+1$ dimensional QCD, where confinement is known to be linear.
In contrast, when one separates the $Q\bar{Q}$ pair transversely,
gauge invariance demands that they are connected by a chain of
(gluon) link fields. For large $a_\perp$, where there are only
little fluctuations, this implies that the energy of such a 
configuration is given by the energy for creating one link
quantum times the number of link quanta, i.e. linear confinement
also in the $\perp$ direction. Since the confinement mechanisms
are very different for these two cases, one might ask whether
a rotationally invariant $Q\bar{Q}$ potential results in the
continuum limit. In fact, in the limit of large $a_\perp$ one
finds in $2+1$ dimensions \cite{conf:prd}:

$V(x_L,x_\perp) = \sigma \left(|x_L|+|x_\perp|\right),$
where $\sigma$ is the string tension, which is clearly not
rotationally invariant.
In order to investigate this issue, I used DLCQ and a Lanzcos
algorithm to calculate
the rest frame $Q\bar{Q}$ potential from the LF Hamiltonian
for $QCD_{2+1}$ on a $\perp$ lattice. The procedure for computing
the rest frame potential in this formalism follows Ref. \cite{zako}.

The light-front Hamiltonian for compact QCD on a transverse
lattice has been introduced by Bardeen \cite{bardeen}. For
pure glue QCD in 2+1 dimensions one finds
\begin{equation}
P^-=c_g \sum_n\int dx^-\int dy^-
:\mbox{tr}\left[ J_n(x^-)\left| x^--y^- \right| J_n(y^-)\right]
:+V_{eff}(U),
\label{eq:pminus}
\end{equation}
where
\begin{equation}
J_n=U^\dagger_n \stackrel{\leftrightarrow}{\partial} U_n
-U^\dagger_{n+1}\stackrel{\leftrightarrow}{\partial} U_{n+1}
\end{equation}
and $U_n$ are the link fields, which are quantized matrix fields and
satisfy the usual
commutation relations. Ideally, one would like to work with
$U_n\in SU(N)$, but in practice this is very complicated \cite{paul:zako}
and it is advantageous to work with an unconstrained complex
matrix field and instead add an effective constraint term
$V_{eff}(U)$ to the light-front Hamiltonian. In the case of
$N\rightarrow \infty$, in the classical limit, 
$V_{eff}(U)$ can be taken of the form
\begin{equation}
V_{eff}^{cl}(U)=c_2 \sum_n\mbox{tr}:\left[U^\dagger_n U_n\right]:
+ c_4 \sum_n\mbox{tr}:\left[ U^\dagger_nU_nU^\dagger_nU_n\right]:,
\label{eq:constr}
\end{equation}
where $c_2=-2c_4$ and $c_4\rightarrow \infty$, which
provides an effective
potential which is minimized for $U_n\in U(N)$. In the $N\rightarrow
\infty$ limit, the difference between $U(N)$ and $SU(N)$ is irrelevant
and Eq.(\ref{eq:constr})
is thus suitable for enforcing the $SU(N)$ constraint in the classical
limit. One might be tempted to try a similar ansatz for the
LF quantized case. There are several reasons why a different form for the
effective potential might be desirable.
First, if one would still attempt to work with an effective
potential of the above form then physical states would
necessarily look extremely complicated, which has to do with the fact
that the above ansatz for the effective potential corresponds to 
working close to the continuum limit. Thus even if the ansatz
in Eq.(\ref{eq:constr}) would work in principle, it would most
likely not be very practical. 
Furthermore, since a "mexican hat"
potential corresponds to a situation where one is working with
the {\it false} vacuum, it is questionable whether a physical
situation where a particle runs at the bottom of a mexican hat
can be described at all by a LF Hamiltonian using degrees of freedom
expanded around the origin.

For these two reasons, it makes more sense not to consider
$U_n$ as the {\it bare} link field, but instead think of it as
some kind of {\it blocked} or {\it smeared} variable. The blocking
has several consequences. First, the $SU(N)$ constraint gets relaxed which
reflects itself in the fact that the effective potential is no longer
just a narrow valley \cite{pirner}. Secondly, using smeared variables,
it might be easier to cover large physical distances with only
few degrees of freedom. 
The price one has to pay for these advantages is that the
effective potential gets more complex and in general more terms are
necessary than shown in Eq.(\ref{eq:constr}). In Ref. \cite{bvds}
an attempt has been made to fit the effective potential to the
glueball spectrum by making an ansatz which includes all operators
up to dimension four.
Since this work is a first study of the rest-frame $Q\bar{Q}$ potential
and, as we will discuss below, the $Q\bar{Q}$ potential turns out
to be rather insensitive to terms of dimension greater than two
in $V_{eff}(U)$, we will instead only consider a much simpler ansatz
in the following and keep only the quadratic term
\begin{equation}
V_{eff}(U) \approx c_2 \mbox{tr}\left(U^\dagger U\right). 
\label{eq:vc2}
\end{equation}
An approximation, where one allows at most one link field
quantum per link, was used. Within this approximation, one obtains
only a first order phase transition at the critical point, i.e.
the lattice spacing always remains finite in physical units.
The calculations were done at the critical point, where
$a_\perp \approx 1.04 \sigma^{-1/2}$ ($a_\perp \approx 1.14 \sigma^{-1/2}$) 
--- with (without) Fock space truncation. 
The resulting $Q\bar{Q}$ potential is shown in Fig. \ref{fig1} as a 
function of $r =\sqrt{x_\perp^2+x_L^2}$.
\begin{Large}
\begin{figure}
\unitlength1.cm
\begin{picture}(1.5,1)(1.35,7.3)
\includegraphics{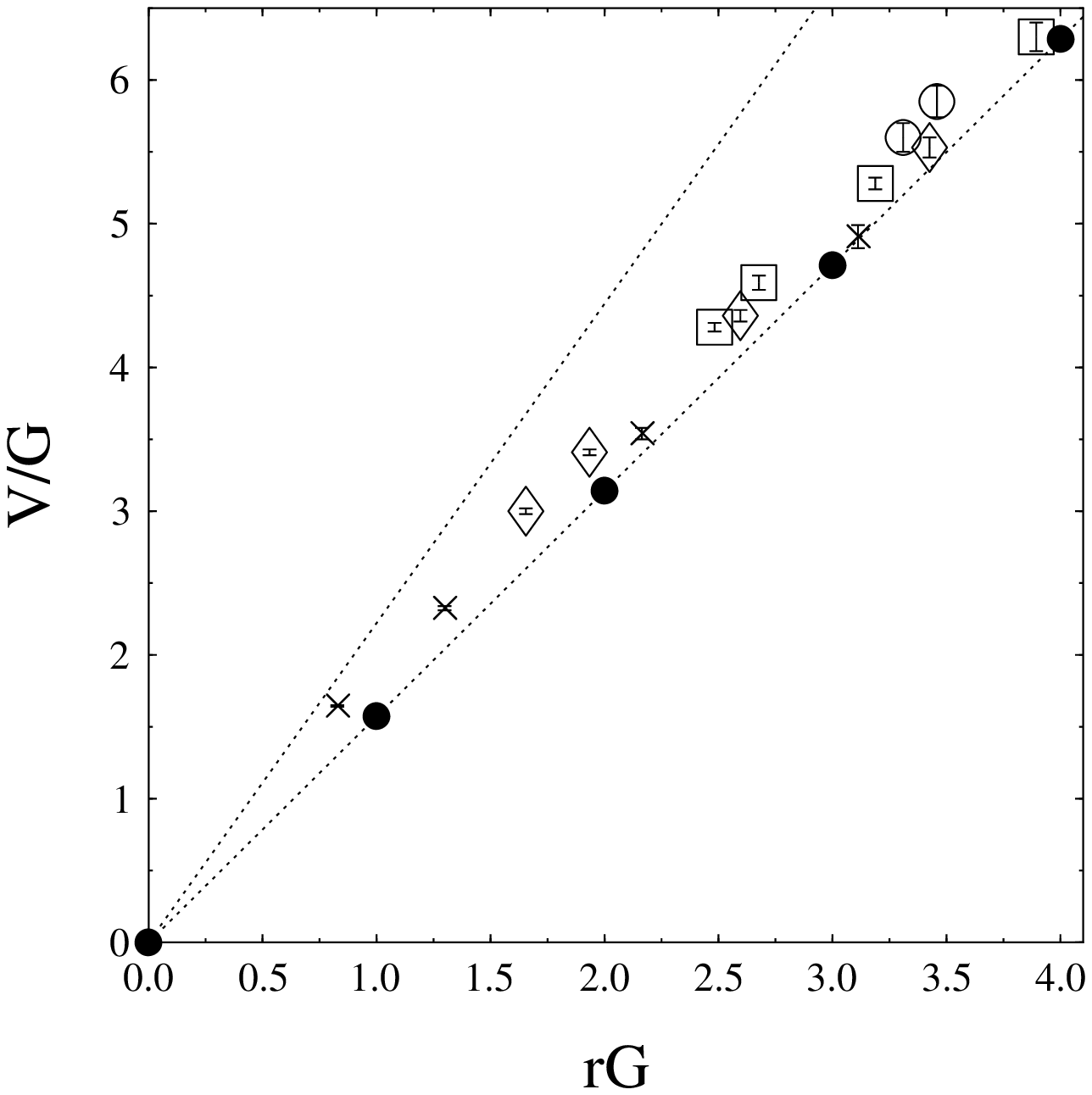}
\put(3.,6.){\makebox(0,0){(a)}}
\end{picture}
\begin{picture}(15,6)(-7,1.3)
\includegraphics{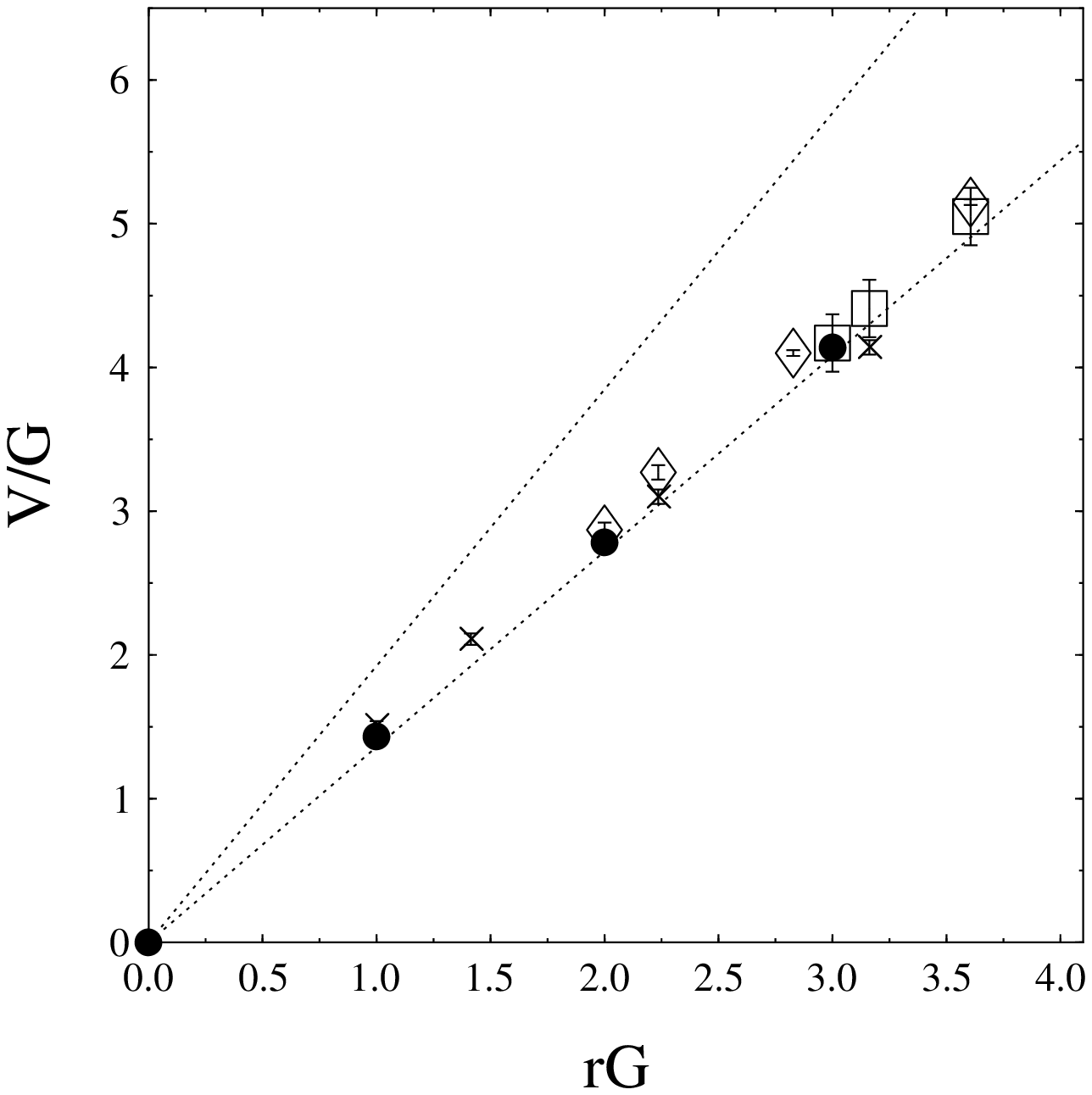}
\put(3.,6.){\makebox(0,0){(b)}}
\end{picture}
\caption{Numerical results for the rest-frame $Q\bar{Q}$ potential
calculated on a $\perp$ lattice, both with (a) and without (b) Fock space
truncation. The various symbols correspond to numerical
results obtained from various directions. All dimensionful quantities are in units of the longitudinal coupling $G$. The dotted lines are extreme cases in the strong coupling limit. The error bars reflect uncertainties in
the continuum extrapolation of the DLCQ results.}
\label{fig1}
\end{figure}
\end{Large}
One free dimensionless parameter was adjusted in the calculation 
to get equal string tensions in the longitudinal
and transverse directions (note that this parameter does not
affect the numerical calculation but is only relevant for scale setting). 
For large $a_\perp$, $V(x_\perp,x_L)$
fills the area between the dotted lines. 
The symbols are numerical results
at the critical point for various directions.
The small residual anisotropy is probably due to the 
use of an oversimplified effective potential.
Note that while QED and QCD are equivalent as long as one restricts
oneself to only one quantum per link (Fig. \ref{fig1}a), this is no longer the case
for the calculations without such a restriction (Fig. \ref{fig1}b).

Intuitively, one would expect a much stronger violation of 
rotational symmetry than in Fig. \ref{fig1}. The reason this did not 
happen is probably the
following: typically, ground states of LF Hamiltonians are
dominated by the lowest Fock component. For the string connecting a
$Q\bar{Q}$ pair, this implies that the wavefunction is dominated by components
with only one quantum per link --- where higher order terms in the
effective potential do not matter. We are currently investigating
excited states of the $Q\bar{Q}$ string, which correspond to
hybrid stated of quarkonium \cite{conf:prd}. In these excited
states, higher order terms in the
effective potential do matter and we expect to obtain information
about the connection between the form of the effective potential 
and rotational symmetry.
\begin{figure}
\begin{Large}
\unitlength1.cm
\begin{picture}(15,8.5)(-1,1.5)
\includegraphics{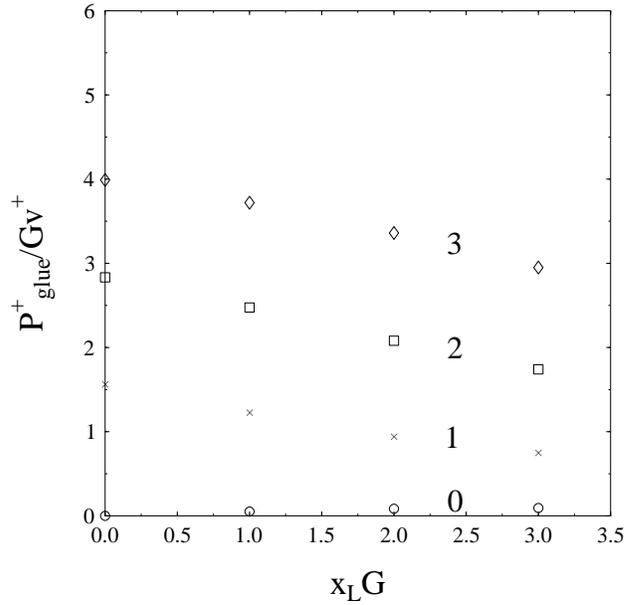}
\end{picture}
\end{Large}
\caption{Momentum carried by the gluons for external charges
that are separated by $n_\perp=0,...,3$ lattice spacings
as a function of the longitudinal (rest-frame) separation of the external
charges. Only the results with inclusion of higher Fock components are shown.
}
\label{fig:xglue}
\end{figure}

Another physical observable for the $Q\bar{Q}$
states is the momentum carried by the glue (Fig. \ref{fig:xglue}).
One finds a strong anisotropy in the sense that
the gluons (link fields) carry much more momentum when the
$Q\bar{Q}$ pair is separated transversely than when it is separated
longitudinally (at the same physical distance, i.e. even though
the same amount of energy is stored in the gluon field in both cases).
Such a result is physical and is familiar from QED \cite{photon}. It can be most
easily understood from the point-of-view of an infinite momentum
boost: if the electric field lines before the boost are $\perp$ to the
boost directions, the field after the boost will look almost like
a transverse electromagnetic wave and carries a large (Pointing) momentum. 
This is not the case when the electric field is parallel to the 
boost direction. 
The anisotropy effect is more pronounced in QCD than in QED because the
(rest frame) electric field is squeezed into flux tubes.
This effect might have measurable consequences in the decays of polarized
hybrid states and shows that the difference between confinement mechanisms in LF-QCD for
longitudinal and $\perp$ directions has physical consequences.


\section*{References}
\begin{thebibliography}{99}
\bibitem{adv} M.~Burkardt, Advances Nucl. Phys. {\bf 23}, 1 (1996).
\bibitem{bardeen} W.~A.Bardeen et al., 
Phys. Rev. D {\bf 21}, 1037 (1980).
\bibitem{conf:prd} M.~Burkardt and B.~Klindworth, submitted to PRD, hep-ph/9601289.
\bibitem{conf:elfe} M.~Burkardt, ELFE Meeting 1995, hep-ph/9510264.
\bibitem{zako} M.~Burkardt, LF-workshop,
Zakopane, Aug 1994, hep-ph/9410219.
\bibitem{paul:zako} P. A. Griffin, LF-workshop,
Zakopane, Aug 1994,
 hep-ph/9410243.
\bibitem{pirner} B. Grossmann et al., {\it Int. J. Mod. Phys.} {\bf A6}, 
2649 (1991). 
\bibitem{bvds} B. vande Sande and S. Dalley, hep-ph/9602291.
\bibitem{photon} M. Burkardt, Nucl.\ Phys.\ {\bf B373}, 613 (1992).
\end{thebibliography}
\end{document}